\documentclass[usegraphicx]{mn2e}
\usepackage{amssymb, amsmath}
\usepackage{ifthen}

\def \version {annotated}
\def \version {clean}

\ifthenelse{\equal{\version}{annotated}}
{
	\usepackage{ulem}			
	\newcommand{\deltext}[1]{\sout{#1}}	
	\newcommand{\newtext}[1]{{\bf #1}}	
	\newcommand{\comment}[1]{{\bf [comment: #1]}}	
	\renewcommand{\em}{\it}
}{
	\newcommand{\deltext}[1]{}		
	\newcommand{\newtext}[1]{{#1}}		
	\newcommand{\comment}[1]{}		
}

\def\br {{\bf r}}
\def\bv {{\bf v}}
\def\bx {{\bf x}}
\def\bg {{\bf g}}
\def\bn {{\bf n}}
\def\ba {{\bf a}}
\def\bp {{\bf p}}
\def\bU {{\bf U}}
\def\bP {{\bf P}}

\def\sc {\rm}

\begin{document}

\title[Redshifts and the Expansion of Space]{Astronomical Redshifts and the Expansion of Space}
\author[Nick Kaiser]{Nick Kaiser \\
Institute for Astronomy, University of Hawaii}
\maketitle 

\begin{abstract}
In homogeneous cosmological models the wavelength $\lambda$ of a photon
exchanged between two fundamental observers changes in proportion to expansion
of the space $D$ between them, so $\Delta\log(\lambda / D) = 0$. 
This is exactly the same as for a pair of observers receding from each other in flat space-time
where the effect is purely kinematic.
The interpretation of this has been the subject of considerable debate, and it
has been suggested the all redshifts are a relative velocity effect, raising
the question of whether the wavelength always stretches in proportion to the
emitter-receiver separation.
Here we show that, for low redshift at least, $\Delta\log(\lambda / D)$ vanishes for a photon exchanged
between any two freely-falling observers in a spatially constant tidal field, because
such a field stretches wavelengths and the space between the observers identically. But in general
there is a non-kinematic, and essentially gravitational, component of the redshift that
is given by a weighted average of the gradient of the tidal field along the photon path.
While the redshift can always be formally expressed using the Doppler formula,
in situations where the gravitational redshift dominates, the `relative velocity' is typically
quite different from the rate of change of $D$ and it is misleading to think of the redshift
as being a velocity or `kinematic'  effect.
\end{abstract}

\begin{keywords}
Cosmology: theory; galaxies: distances and redshifts
\end{keywords}

\comment{This is marked up version to show the changes from submitted version.  To generate clean
version uncomment the line that defines `version' as `clean' near top of latex source file}

\section{Introduction}

\subsection{Overview, Goals and Outline}

\comment{Intro section now split into two subsections as per reviewer's comments.}

In spatially homogeneous and isotropic Friedmann, Robertson \& Walker (FRW) cosmological models
the wavelength $\lambda$ of a photon exchanged between any two
fundamental (or `co-moving') observers (FOs) changes in proportion to the change in the scale factor $a(t)$
or, equivalently, in proportion to the change in the proper separation $D$ of the
observers, which for concreteness we take to be the integrated distance along the geodesic
of the 3-space of constant proper time since the big-bang that connects them, so
\begin{equation}
\Delta\log(\lambda / D) = 0.
\label{eq:stretchingrelation}
\end{equation}  
This `cosmic wavelength stretching relation' also applies for FOs in
homogeneous but anisotropic models.  This is a well-known and familiar result.
The stretching is sometimes described as being caused by the expansion of space,
and can also be readily understood in terms of standing waves in an expanding cavity.
But is also perhaps a little
surprising since it has been known since Bondi (1947) that the redshift between FOs
is, for low redshift at least, expressible as the product of a 
special relativistic (SR) Doppler shift and a gravitational redshift,
these terms generating fractional wavelength shifts that are 
respectively linear and quadratic in distance (see Peacock 2008).
Yet the stretching relation (\ref{eq:stretchingrelation}) is just the
same as for a pair of observers receding from one another in flat space-time; the
effect of gravity does not appear.   

Subsequently, Synge (1960) showed
that {\em any\/} redshift is expressible as a Doppler
shift, with the relative velocity defined in terms of parallel transport of the
emitter's 4-velocity.  This provides an elegant and unified way to view all redshifts.
Significantly, Synge says that redshifts should not be considered to be
a gravitational effect as the curvature tensor does not appear in the formulae.

This view has been revived by Narlikar (1994) and, more recently, by Bunn \& Hogg (2009)
who argue that all redshifts are Doppler, or `kinematic', in nature because, in their view, 
a gravitational redshift is just a Doppler shift viewed from an unnatural coordinate system.

Our interest in this was piqued by recent measurements of gravitational redshifts
from clusters of galaxies (Wojtak et al.\ 2011).  They measured the
potential well depth difference between low specific energy brightest cluster galaxies (BCGs) and the general
cluster population.  Theoretical studies
(e.g.\ Cappi 1995) considered the redshift $\Delta \lambda / \lambda = \Delta \phi / c^2$ that would be seen
by static non-inertial observers,
with $\Delta \phi$ the Newtonian gravitational potential difference.  
But the sources and observer are really in free fall.  Does this make any difference?
Obviously the main effect of this is to add a very large first order Doppler
effect from the relative motion, which has to be carefully averaged out to reveal the
gravitational effect.  Another complication is that one will see transverse Doppler
and other kinematic effects that are generally of the same order as the
static gravitational redshift (Zhao et al.\ \deltext{2012} \newtext{2013}; Kaiser 2013).
But imagine, if you will, a freely-falling source and observer pair 
that both reside in the potential well of a relaxed static massive cluster and whose separation
is known to have been the same at the times of emission and reception.  What redshift would they see?

It would seem natural to interpret the kinematic view as saying that there
would not be any redshift.  After all this would be the case for a pair of FOs inside a small part of a 
closed FRW model at its phase of maximum expansion. Is this fundamentally any different from the observers in the cluster? 

Another possibility would be to argue that they would not see
any first order potential difference  $\Delta \phi \sim - {\bf D} \cdot \bg$
from any spatially constant component of $\bg$ by virtue of the equivalence principle, but perhaps
one might expect them to perceive any variation of the gravity, i.e.\ to lowest order at small separation
the local tidal field.  But there are some problems with this view.  One is that
there is a non vanishing tide in the closed FRW example cited above, yet apparently
this does not give rise to a redshift.  
Also, in this picture, any observer in a smooth cluster potential may consider itself to be at the bottom of a locally 
parabolic (though possibly tri-axial) potential well.  This is certainly legitimate in the sense that,
if that observer \deltext{were} were to release a test particle at rest nearby, it would indeed see
that particle accelerate slowly towards it.   That might lead one to
think that all observers would see an average blue-shift for photons from neighbour particles
that are, on average, not moving away or approaching the chosen observer.

But both of these might seem to conflict with the view -- arguably the prevailing one in the field
of cluster gravitational redshifts -- that if the emitters are BCGs on low energy orbits close to the
centre of the cluster and the receivers are galaxies on radial orbits much further out
then they actually {\em would\/} see the full potential difference $\Delta \phi$ including any
first order $\sim - {\bf D} \cdot \bg$ effect as well as the higher order terms.

The goal of this paper is to try to see how the seemingly different, and perhaps conflicting, views 
described above can all be reconciled in a coherent way and to develop a consistent picture for
how to think about redshifts in general, and particularly  gravitational redshifts, in a way that avoids
misconceptions or pitfalls.  

The outline of the paper is as follows: In the rest of this Introduction we review the history of how the
homogeneous cosmology wavelength stretching has been interpreted and also review the extension to 
inhomogeneous situations.  As our results are closely related to the the proposal of Bunn \& Hogg
we review their arguments in some detail.
In Section 2 we perform a simple calculation of the
domain of validity of the stretching relation by directly calculating the change in wavelength and
proper separation of observers in arbitrary gravitation fields, though the calculation is
limited to weak fields (i.e.\ low redshifts).  We conclude with a summary and discussion, including an illustrative
example.

\subsection{Different Interpretations of the Redshift}

In many popular accounts the relation $\Delta\log(\lambda / D) = 0$
is described as a causal effect of the expansion of space
stretching the wavelength of light.  
Aside from giving the right answer, this finds support in the fact that Maxwell's equations written
in the co-moving coordinates that are most natural in an expanding universe have an extra `damping' or `friction'
term that gives a reduction of co-moving energy density -- consistent with occupation numbers being adiabatically
conserved while wavelengths get stretched (reducing the energy per photon) -- and that these equations admit e.g.\ standing waves where
the separation of nodes grows with the expansion factor.  In this picture, the red-shifting of all radiation as $\lambda \propto a(t)$
appears inescapable and it is often stated as if self-evident that if the scale factor doubles the wavelength of light must double too
(e.g.\ Harrison, 2000; Lineweaver \& Davis, 2004).
  
These damping standing wave solutions are a good way to think about the cosmic microwave background, 
and this line of argument has the virtue of giving the right answer for photons exchanged between FOs also.
But the `expanding-space' paradigm has been criticised, mostly on the grounds that it obscures
the central tenet of general relativity (GR), which is that space-time is 
locally flat, or Minkowskian (e.g.\ Whiting 2004; Peacock 2008; Bunn \& Hogg 2009; Chodorowski 2007, 2011), so 
it does not define any local state of expansion.  Writing Maxwell's equations in expanding coordinates does not change
the physics,  or the solutions, and nothing in the physical laws says that wavelengths should increase.
This problem is seen most starkly in matter-free FRW models such as in de Sitter space-time
and the  Milne model in Minkowski space-time which admit solutions for expanding (or contracting) families of FOs defining different
expansion histories $a(t)$ -- possibly overlapping in the same region of space-time -- and also
admit solutions to Maxwell's equations that correspond to either red-shifting or blue-shifting radiation fields.
In these solutions, the expansion rate is defined by the radiation itself;
an observer who is in the zero momentum density frame at some point will see a Poynting flux
that increases linearly with distance, and other observers who locally see zero momentum flux
obey Hubble's law.  This is all ultimately determined by the initial condition for the field. 
Additionally, it may not be completely obvious how e.g.\ a classical wave packet being
emitted and received by localised observers relates to an unbounded, perhaps infinite, standing wave.

The most satisfactory explanation of the wavelength-separation relation in FRW models
is that of Peebles (1971). He argued that the overall wavelength ratio
for a widely separated pair of FOs is the product of the ratios for a set of intervening
exchanges between neighbouring FOs along the photon path, each of which are, to first order, Doppler effects because of the
local flatness of space-time.  This is clearly `GR compliant'  and leads to the differential equation $d \lambda / \lambda = d a / a$
with solution $\lambda \propto a(t)$.  In this picture the change in wavelength can be thought of as an unchanging
photon being seen by a succession of observers with different frames of reference.

While the interpretation of the cosmic wavelength stretching has been contentious, 
no one doubts that it is obeyed in homogeneous models.  The question we shall ask here is whether this
is of broader validity and whether it is valid in an inhomogeneous universe.
It is not clear how the expanding space paradigm can be extended to inhomogeneous gravitating
systems.  
One might try to make progress by
modelling single-flow regions (if such exist) as being locally like anisotropic
homogeneous models, but since galaxies and dark-matter may, in general, have multiple
streaming velocities at the same point in space it is obvious that one cannot even
think about the local space in a bound system like a cluster of galaxies as having
any state of expansion.  But the concept of the space between any pair of observers or
galaxies changing with time is still perfectly valid,  and so one can
ask, for instance, whether (\ref{eq:stretchingrelation}) still applies.

Relaxing the requirement of homogeneity,
Bondi (1947) showed that, for small redshifts at least and in matter dominated spherically symmetric models, the
redshift can be expressed as the product of a Doppler shift and a gravitational redshift (see Peacock 2008).  
This might be taken to suggest that there is something fundamentally different
about inhomogeneous models.  
But this applies also to the homogeneous case, and is not inconsistent with the wavelength
obeying the `stretching law' (\ref{eq:stretchingrelation}).  The resolution of the apparent difference is that Bondi's velocity is
that of the receiver relative to the emitter at the time of reception, with the emitter being stationary
at the centre of the expanding sphere of matter on whose surface the receiver resides (Peacock, 2008).  
This is not what determines the change in the proper separation, which is some average
of the observers relative velocity over the light travel time and it is not hard to show that allowing for the
change of this velocity is equivalent to including the extra gravitational redshift factor.

Relaxing the requirement of spherical symmetry and matter domination, Synge (1960) has shown that the 
redshift is quite generally given by the usual special relativistic Doppler formula alone, with
the `relative velocity' being defined in terms of 
parallel transport of  the emitter 4-velocity along the null ray from the emission event
to the reception event.  The energy perceived by an observer (emitter or
receiver) is the dot product of their 4-velocity $\bU$ with the photon 4-momentum $\bP$.
These observed energies transform as scalars, and so can be directly compared, but this comparison
can also be made by parallel transporting 
the emitter velocity and photon momentum from the emitter to the receiver (along the photon's null path)
and then comparing.  But transporting the photon along its path does not change it, so the result,
with S and O denoting source and observer, is
\begin{equation}
\frac{\lambda_{\rm 0}}{\lambda_{\rm S}} = \frac{\bU_{\rm S}\cdot\bP_{\rm S}}{\bU_{\rm O}\cdot\bP_{\rm O}}
= \frac{{\tilde\bU}_{\rm S}\cdot\bP_{\rm O}}{\bU_{\rm O}\cdot\bP_{\rm O}}
\label{eq:syngelambdaratio}
\end{equation}
where  ${\tilde\bU}_{\rm S}$ is the parallel transported source 4-velocity.
The last expression above is the usual SR Doppler effect (since 
all of the 4-vectors are defined at the point of reception and
at this point, as at any other, space-time is locally flat).  

This is an elegant  formalism that nicely unifies all redshifts, be they Doppler,
gravitational, or cosmological (Narlikar 1994).
However, in discussing this, Synge says that the spectral shift
is a velocity effect, and not a gravitational effect, because the Riemann curvature tensor
does not appear in (\ref{eq:syngelambdaratio}).  This is surprising, since
the parallel transport $\bU_{\rm S} \rightarrow {\tilde \bU}_{\rm S}$ depends on the
connection, which, while not the same as the curvature is closely related to it.

Regarding the physical interpretation of Synge's velocity,  Chodorowski (2011)
showed that, in FRW models, the difference between parallel transporting from the emitter to the
receiver along the photon's null-path and parallel transporting along the geodesic
lying in the 3-space of constant proper time (at the time of emission, say) is
just Bondi's gravitational component of the redshift.
This can also be understood in the weak-field \deltext{\&} \newtext{and} non-relativistic observer limit, which is applicable within
a limited region of an FRW model.  Parallel transport in time alone is then a rotation of the 4-velocity $U$
at \deltext{a} \newtext{an} angular frequency $g/c$, \newtext{where $g$ is the Newtonian gravity,} 
so for non-relativistic observers:  $dU/dt =  \{0, \bg/c\}U^0$.   In FRW models, the gravity $\bg$ 
increases linearly with distance from any reference position, so $\bg = - \br d^2 \phi / d r^2$.  
The photon time of flight is $\Delta t = r / c$, so $\delta v = (r/c) dU_r/dt = - (d^2 \phi / d r^2) r^2 / c$,
which is quadratic in separation and proportional to the constant tidal field.




More recently,  Bunn \& Hogg (2009), in their stimulating critique
of the expanding space picture, discuss redshifts in the more general context of
non-homogenous universes.  
Like Peebles, they consider the overall wavelength ratio to be the product of incremental
shifts between neighbouring observers along the photon path and invoke local flatness.
They argue that:
\begin{enumerate}
\item An observed frequency shift in
any space-time can be interpreted as either a kinematic (Doppler) shift or
a gravitational  shift by imagining a suitable family of observers along the photon's path.
\item It is more natural to consider them to be Doppler or kinematical in nature.
\item In situations where any relative velocities are $\ll  c$,
and curvature is small (over the distance and time scales traveled
by the photon) one should have no hesitation in applying the Doppler formula.
\end{enumerate}

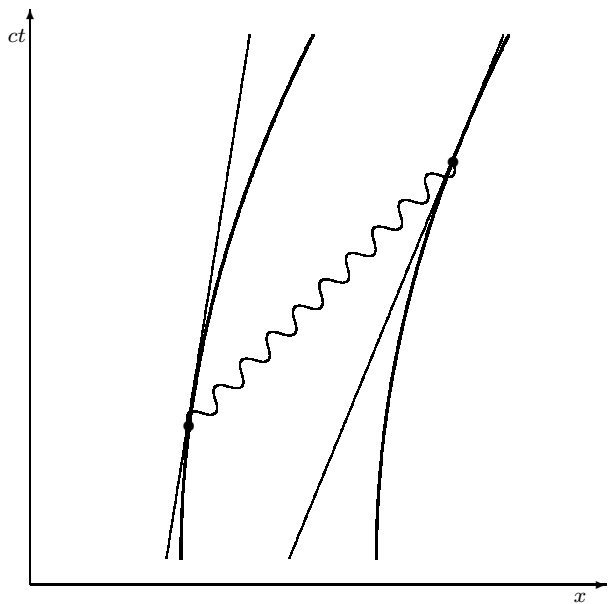
\begin{figure}
\begin{picture}(238,238)(0,0)
\thinlines
\put(10,10){\vector(1,0){218}}
\put(218,5){\makebox(0,0){$x$}}
\put(10,10){\vector(0,1){218}}
\put(5,218){\makebox(0,0){$ct$}}
\qbezier(  70.00,  70.00)(  67.50,  77.50)(  75.00,  75.00)
\qbezier(  75.00,  75.00)(  82.50,  72.50)(  80.00,  80.00)
\qbezier(  80.00,  80.00)(  77.50,  87.50)(  85.00,  85.00)
\qbezier(  85.00,  85.00)(  92.50,  82.50)(  90.00,  90.00)
\qbezier(  90.00,  90.00)(  87.50,  97.50)(  95.00,  95.00)
\qbezier(  95.00,  95.00)( 102.50,  92.50)( 100.00, 100.00)
\qbezier( 100.00, 100.00)(  97.50, 107.50)( 105.00, 105.00)
\qbezier( 105.00, 105.00)( 112.50, 102.50)( 110.00, 110.00)
\qbezier( 110.00, 110.00)( 107.50, 117.50)( 115.00, 115.00)
\qbezier( 115.00, 115.00)( 122.50, 112.50)( 120.00, 120.00)
\qbezier( 120.00, 120.00)( 117.50, 127.50)( 125.00, 125.00)
\qbezier( 125.00, 125.00)( 132.50, 122.50)( 130.00, 130.00)
\qbezier( 130.00, 130.00)( 127.50, 137.50)( 135.00, 135.00)
\qbezier( 135.00, 135.00)( 142.50, 132.50)( 140.00, 140.00)
\qbezier( 140.00, 140.00)( 137.50, 147.50)( 145.00, 145.00)
\qbezier( 145.00, 145.00)( 152.50, 142.50)( 150.00, 150.00)
\qbezier( 150.00, 150.00)( 147.50, 157.50)( 155.00, 155.00)
\qbezier( 155.00, 155.00)( 162.50, 152.50)( 160.00, 160.00)
\qbezier( 160.00, 160.00)( 157.50, 167.50)( 165.00, 165.00)
\qbezier( 165.00, 165.00)( 172.50, 162.50)( 170.00, 170.00)

\put(70.000000,70.000000){\circle*{4}}
\put(170.000000,170.000000){\circle*{4}}
\thicklines
\qbezier(67.000000,20.000000)(67.000000,119.000000)(117.000000,218.000000)
\qbezier(141.000000,20.000000)(141.000000,119.000000)(191.000000,218.000000)
\thinlines
\qbezier(61.500000,20.000000)(77.250000,119.000000)(93.000000,218.000000)
\qbezier(108.000000,20.000000)(148.500000,119.000000)(189.000000,218.000000)
\end{picture}
\caption{Schematic space-time diagram for exchange of a photon 
in flat space-time between a pair of freely-falling observers
(thin lines) and between a pair of observers being subject to non-gravitational acceleration (thick lines).  
Relative motions and accelerations are assumed here to be
aligned with the photon path.  Bunn \& Hogg (2009) pointed out
that for any such photon path and freely falling observers the emission and reception events
can be taken to lie on the world lines of a pair of observers
who live on opposite ends of a
uniformly accelerating rod with those world lines being tangent to
those of the freely falling observers.  This is possible since one can
choose the initial position and velocity
of the rod to make the observer at one end of the rod be co-located
and co-moving with the freely falling emitter at the emission event and
one can then choose the length of the rod and its acceleration so that,
by the time the photon reaches the freely falling receiver the other
end of the rod has caught up with it. 
The accelerated observers perceive the rod to have fixed length,
though in the `lab-frame' the rod will appear progressively foreshortened. 
The freely falling observers
view the redshift as a Doppler effect with $\Delta \lambda / \lambda = \Delta v / c$ (for $\Delta v \ll c$)
caused by their relative motion.   The accelerated observers would note
that the redshift is related to their acceleration $a$ and the rod length $l$
by $\Delta \lambda / \lambda = a l / c^2$. \comment{t axis changed to ct}}
\label{fig:poundrebka}
\end{figure}

The latter two statements resonate with Synge's,
but the way that are reached is actually different.
The logic used here is that, while it might seem most natural to
consider the imaginary intervening observers to be freely falling, one
could also imagine them to be accelerated observers, with rocket motors
strapped to their legs perhaps.  If they are each co-moving with one of another
family of freely-falling observers at the moment the photon passes, the
same incremental wavelength shifts would apply since all
that matters for the redshift is the tangent to the
4-velocity at the events of emission and reception, any curvature
of the world lines before or after the event being irrelevant. 

They then invoke the `parable of the speeding ticket' in which a motorist 
caught speeding by a cop with a radar gun 
tries to avoid the fine by arguing that in a different coordinate frame the car
and the cop were not moving relative to each other (though perhaps a more
precise analogy would be a cop measuring the frequency of
radiation from a source of known frequency mounted on the car).
As illustrated in Fig.\  \ref{fig:poundrebka}, one can indeed imagine a rod
being uniformly accelerated (by rocket motors, say) passing by with
two observers riding on it, one being present at the emission
event and co-moving with the car and the other present at the reception
instantaneously co-moving with the cop.
In the coordinate system of the accelerated rod, it is claimed,
the emitter and receiver are not in a state of relative motion; so there
is no Doppler shift, but there is now a `gravitational' redshift $\Delta \lambda / \lambda = a l / c^2$
where $a$ is the acceleration and $l$ is the length of the rod.
Thus, they argue, an enlightened cosmologist would never try to
characterise a redshift as a velocity or gravitational effect, because the
two interpretations arise from different choices of coordinates.

This is a fascinating and ingenious argument, but probably not one that Synge would have approved of.
As he was at pains to emphasise in the preface to his book, despite its name, GR is an absolute theory
since whether or not there is a gravitational field in some region of space is unambiguously
measurable from geodesic deviation of freely-falling test particles (though the 
values of the components of the curvature tensor are coordinate system dependent). 
The curvature, or tidal field, is unaffected by the presence of any observers (real or imaginary) who might be
accelerated by rockets.\footnote{Rindler (1970) gives an interesting argument, which
he attributes to Dennis Sciama, that the weight of objects sensed by an accelerated
observer in a rocket can be thought of, in a Machian sense, as gravity arising from the
relative acceleration of the rest of the Universe.  That argument cannot be applied
here, since the acceleration of the imaginary intervening observers is determined
by the arbitrary choice of their velocities; this is generally varying along the photon
path and the gradient of this is not equal to the real tidal field.}
If the curvature vanishes in the region of space-time containing the
observers and the photon path then whatever happens there
can hardly be said to be a gravitational redshift.


Similarly, while the velocity of an object depends on the
frame from which it is observed, the relative velocity of
two objects in their centre of velocity frame is another absolute quantity.
Accelerated observers know that they are being accelerated.
Once they allow for this the accelerated observers here would be in
full agreement with the cop as to how fast the
motorist was approaching.

It is true that in the Pound \& Rebka (1959) experiment the wavelength shift $\Delta \lambda / \lambda = gh/c^2$ 
they measured is the same as
the (constant) relative velocity of a pair of hypothetical freely-falling observers launched so as to
be tangent to the world-lines of the actual emitter and receiver at the interaction events (this
being the relative velocity in the `lab' or in the centre of velocity frame -- the difference
being negligible -- but not the difference in velocities at times of the actual events).  But that is
just telling us that this experimental result is fully accounted for by the fact that the
real apparatus is being accelerated by {\em non-gravitational\/} 
stresses in the instrument supports and in the  planet that is standing in the way of its natural free fall.
From a Syngean perspective, Pound \& Rebka did not measure a {\em gravitational\/} redshift at all
as their experiment was simply not sensitive enough to measure the gravitational curvature or tide.


Accelerated observers are interesting, but are something of a distraction.
For redshifts between galaxies there are no non-gravitational forces to worry about; 
all real sources and observers are freely-falling. 
Knowledge of the tidal field in the vicinity of the observers and along the photon path is then
all that is needed to calculate how the observers' motions
evolve and how photons exchanged between them get redshifted. 
It does not matter that the {\em gravity\/} $\bg$ is only determined by local measurements
up to an additive constant vector 
as  that has no effect on any measurements
made by observers in free-fall in the region where the tide has been determined.\footnote{For example, 
while it is widely believed that the dipole anisotropy of the
microwave background is the result of our being accelerated by large-scale structure,
it is possible that some of the dipole is generated by a large-scale 
specific entropy gradient (Gunn 1988), but this indeterminacy of the local value of $\bg$ has no effect on local dynamics
within the milky way or within the local supercluster say.}

So there is no ambiguity
in defining the gravitational field, or in calculating its influence on photons or observers' trajectories.
The only possible ambiguity here is that if there is non-vanishing 
tidal field and if one tries to decompose the redshift into a 1st order Doppler effect and a gravitational
effect then the latter will depend,
possibly quite sensitively, on the time at which one choses to compare velocities to obtain the first order term.  
This is analogous to the interpretation of the Bondi gravitational term as a correction of the
Doppler term from final time to average time (see also Chodorowski 2011).  But the redshift itself is not ambiguous, and if the relative
velocity is chosen to be either at the time of emission, reception or, say, half way along the photon path
there is no ambiguity.  And, as we shall see, if we compare the redshift to the change in
separation $D$ -- which involves an average of the velocity over the photon travel time -- there is no ambiguity either.

What then of points (ii) and (iii)
above, that one should consider all redshifts to be Doppler, or kinematic, in nature
and that, perhaps with restrictions on recession velocity and curvature, one should be confident in
applying the Doppler formula?

The legitimacy of considering any redshift to be the product of a lot of little
Doppler shifts (with the understanding that the relative velocity for each pair of particles is in their
centre-of-velocity frame) is not in question.  Nor is Synge's result that the overall redshift is given by the Doppler
formula and that, for low redshift, Synge's velocity must be equal to the sum of the incremental velocities
as defined here.  But the burning question is: What does this net `velocity' mean physically?
As Synge said, arguments about whether a spectral shift is a gravitational or velocity
effect are just `windy warfare' without analysing the meaning of the terms being employed.
So, how {\em does\/} Synge's velocity relate to the actual relative velocity of the source and observer?

Bunn \& Hogg shy away from this question and deny the legitimacy of considering the relative velocity or
separation of observers as we have used the terms above.
They argue that one needs to talk about $v_{\rm rel}$, the
velocity of a galaxy {\em then\/} relative to us {\em now\/};
that it is hard to define relative velocity of two separated observers in curved space-time;
that all one can do is parallel transport,
and that the only `natural' choice of path is the null-ray of the photon.  
That all sounds very reasonable.
But if this were the only way to define relative velocity then
to say that redshift is a velocity effect would be circular.  Saying that
the redshift is given by the Doppler formula if the velocity can only
be determined by taking the inverse Doppler function of the redshift is not
very useful.  In contrast, saying that the redshift between FOs in FRW models
is kinematic is meaningful since the fractional change of wavelength
really is equal to the fractional change in proper separation.  This is not 
a mathematical identity, but a relationship that holds between
two distinct physical entities. 


There are many ways, in principle at least, to directly measure the relative
velocity of a pair of observers that are independent of the redshift, at least at low redshift.  
One can use rate of change of parallaxes or
light-echo time delays.  It has been suggested
that one can directly measure velocities on cosmological scales
by the shrinking of the angular size of bound objects with time
(Darling 2013).  And one could, in principle, use rate of change of
luminosity of standard candles. 
It may not be practically achievable, but one can imagine a very large rigid
lattice populated with observers with clocks and rulers
who record the rate of motions of observers flying past. 
This, after all, is how we usually imagine measuring
geodesic deviation in order to determine the gravitational field.
The question here is not whether
this can be done in practice; only whether it is possible in principle.
As we describe below, we are free to construct the
lattice such that the velocities of the emitter and receiver relative
to the lattice are equal and opposite.  All of these concepts converge,
at low redshift, to an
unambiguous operational definition of relative velocity in the centre of velocity 
frame.\footnote{It is hard to understand
the reluctance to consider such velocities as legitimate; if there are
any problems with the concept of relative velocity of two observers
at the same {\em time\/} (i.e.\ in their centre of 
velocity frame) they seem to us to pale into insignificance compared to the more
fundamental problem of defining the difference of velocities at
different times in the face of the unknown absolute value of $\bg$.}
This, of course, is the physical quantity that one would normally
associate with the phrase `relative-velocity' (rather than the relatively abstract definition in
terms of the mathematical operation of parallel transport). 

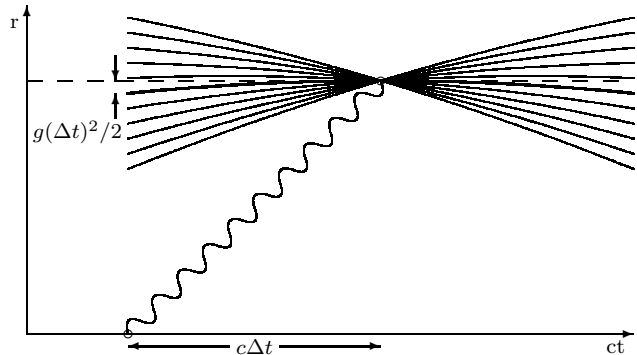
\begin{figure}
\setlength{\unitlength}{3.360000cm}
\begin{picture}(2.5,1.4)(-1.5,-1.1)
\put(-1.400000,-1.000000){\vector(1,0){2.400000}}
\put(0.930000,-1.050000){\makebox(0,0){ct}}
\put(-1.400000,-1.000000){\vector(0,1){1.300000}}
\put(-1.450000,0.230000){\makebox(0,0){r}}
\multiput(-1.400000,0.000000)(0.120000,0.000000){20}{\line(1,0){0.060000}}
\put(-1.0,-1.0){\circle{0.03}}
\put(0.0,0.0){\circle{0.03}}
\qbezier(  -1.00,  -1.00)(  -1.02,  -0.92)(  -0.95,  -0.95)
\qbezier(  -0.95,  -0.95)(  -0.87,  -0.98)(  -0.90,  -0.90)
\qbezier(  -0.90,  -0.90)(  -0.92,  -0.83)(  -0.85,  -0.85)
\qbezier(  -0.85,  -0.85)(  -0.77,  -0.88)(  -0.80,  -0.80)
\qbezier(  -0.80,  -0.80)(  -0.82,  -0.73)(  -0.75,  -0.75)
\qbezier(  -0.75,  -0.75)(  -0.67,  -0.78)(  -0.70,  -0.70)
\qbezier(  -0.70,  -0.70)(  -0.72,  -0.63)(  -0.65,  -0.65)
\qbezier(  -0.65,  -0.65)(  -0.57,  -0.68)(  -0.60,  -0.60)
\qbezier(  -0.60,  -0.60)(  -0.62,  -0.53)(  -0.55,  -0.55)
\qbezier(  -0.55,  -0.55)(  -0.47,  -0.58)(  -0.50,  -0.50)
\qbezier(  -0.50,  -0.50)(  -0.52,  -0.43)(  -0.45,  -0.45)
\qbezier(  -0.45,  -0.45)(  -0.37,  -0.48)(  -0.40,  -0.40)
\qbezier(  -0.40,  -0.40)(  -0.42,  -0.33)(  -0.35,  -0.35)
\qbezier(  -0.35,  -0.35)(  -0.27,  -0.38)(  -0.30,  -0.30)
\qbezier(  -0.30,  -0.30)(  -0.32,  -0.23)(  -0.25,  -0.25)
\qbezier(  -0.25,  -0.25)(  -0.17,  -0.28)(  -0.20,  -0.20)
\qbezier(  -0.20,  -0.20)(  -0.22,  -0.13)(  -0.15,  -0.15)
\qbezier(  -0.15,  -0.15)(  -0.07,  -0.18)(  -0.10,  -0.10)
\qbezier(  -0.10,  -0.10)(  -0.12,  -0.03)(  -0.05,  -0.05)
\qbezier(  -0.05,  -0.05)(   0.03,  -0.08)(   0.00,  -0.00)

\qbezier(-1.000000,-0.348000)(0.000000,0.052000)(1.000000,0.252000)
\qbezier(-1.000000,-0.288000)(0.000000,0.052000)(1.000000,0.192000)
\qbezier(-1.000000,-0.228000)(0.000000,0.052000)(1.000000,0.132000)
\qbezier(-1.000000,-0.168000)(0.000000,0.052000)(1.000000,0.072000)
\qbezier(-1.000000,-0.108000)(0.000000,0.052000)(1.000000,0.012000)
\thicklines
\qbezier(-1.000000,-0.048000)(0.000000,0.052000)(1.000000,-0.048000)
\thinlines
\qbezier(-1.000000,0.012000)(0.000000,0.052000)(1.000000,-0.108000)
\qbezier(-1.000000,0.072000)(0.000000,0.052000)(1.000000,-0.168000)
\qbezier(-1.000000,0.132000)(0.000000,0.052000)(1.000000,-0.228000)
\qbezier(-1.000000,0.192000)(0.000000,0.052000)(1.000000,-0.288000)
\qbezier(-1.000000,0.252000)(0.000000,0.052000)(1.000000,-0.348000)
\put(-1.050000,0.100000){\vector(0,-1){0.100000}}
\put(-1.050000,-0.148000){\vector(0,1){0.100000}}
\put(-1.200000,-0.200000){\makebox(0,0){$g (\Delta t)^2 / 2$}}
\put(-0.500000,-1.050000){\makebox(0,0){$c\Delta t$}}
\put(-0.600000,-1.050000){\vector(-1,0){0.400000}}
\put(-0.400000,-1.050000){\vector(1,0){0.400000}}
\end{picture}
\caption{Illustration of situation described in text where a photon
is emitted by a `cold' particles near the centre of a smooth potential well and
is received by a randomly chosen `hot' particle at large distance.  A sample of orbits
from the distribution of velocities -- here a simple box-car -- is shown as curves.
The orbit for the average velocity particle is shown as the heavy curve.
The average radial velocity is zero at the time of reception, but the
average velocity over the time-of-flight of the photon is $-g \Delta t / 2$
and, for a parabolic potential, this velocity, in units of $c$ is equal to the
gravitational redshift.
It is tantalising to think that a generalisation of this reconciles the
GR kinematic view of redshifts with more conventional view of
gravitational redshifts.
\newtext{As we will see, however, this result only obtains for this specific,
rather special, form for the potential and does not apply in the general case.}
\comment{t axis label changed to ct}}
\label{fig:clusterfig}
\end{figure}

In the weak-field limit, and for stationary non-inertial observers,
Synge's relative velocity is $\Delta v = \int d\br \cdot \bg / c$.  As discussed,
in the context of the Pound-Rebka experiment, this is the same as the constant relative velocity of
a pair of particles launched so as to be tangent at the interaction events. But that is a flat-space (constant
gravity) phenomenon.
More interesting is how the velocities are related when curvature cannot be neglected.
To this end, consider test particles in dynamic equilibrium in a static potential
well.  Imagine a dark matter halo that generates a smooth bowl-shaped potential well
and imagine two families of test particles; a `hot' high energy population and
a `cold' low energy population that have have relatively negligible velocities and
are confined to the very centre of the halo.  The conventional view is that photons emitted from
a cold particle and received by a randomly chosen hot particle at larger radius
will on average suffer the usual static gravitational redshift $\langle \Delta \lambda \rangle / \lambda \simeq \Delta \phi / c^2$,
and that photons emitted by a randomly chosen hot particle and received by a cold particle
will be blue-shifted by the same amount.  This is the redshift
that which would be seen by a static non-inertial observer being supported against falling by stress in
its supporting structure, but the hot population is similarly being supported against falling by the stress of 
its kinetic pressure so it should be essentially identical.  Though as mentioned above, the redshift is not exactly the same because there will also
be a transverse Doppler and other second order kinematic effects
that give corrections to the naive prediction that, by virtue of the virial theorem, is guaranteed to be of the
same order of magnitude (though will in general dependent on the details of the strucure and velocity dispersion of the cluster).  

In this example we see \deltext{the} the important, and general, distinction between gravitational and Doppler
redshifts; that gravitational redshifts are generally asymmetric and change sign if we reverse the direction
of light propagation, while Doppler shifts are symmetric; both observers
perceive a shift of the same sign.  This means that Synge's parallel transport
velocity difference here is, in an average sense, positive in the first case (out-going photons) and negative in the latter, whereas 
randomly chosen hot particles have, of course, zero real average relative velocity.  

On the face of it this makes the kinematic picture seem a bit nonsensical; how can the relative
velocity of the two populations depend on who was emitting and who was receiving? 
But a moment's reflection reveals that in the case of out-going (in-going) photons, the
hot halo particles that interacted with the photons actually did have a positive (negative) average radial velocity  {\em during the
time of flight of the photon\/}, as illustrated in Fig.\ \ref{fig:clusterfig}.    Consider, for simplicity, a parabolic potential
well: $\phi = \alpha r^2$.  The average velocity, over the time of flight for an out-going photon,
is $\langle v \rangle = -g \Delta t / 2 = \alpha r^2 / c$, so $\langle v \rangle / c = \langle \Delta \lambda \rangle / \lambda$,
just the usual gravitational redshift.  So Synge's velocity can here be interpreted as the
pair-wise velocity averaged over the light travel time.
Moreover, the change in separation during the 
photon trip is $\langle \Delta D \rangle  = g (\Delta t)^2 / 2$, so, since $\Delta t = D/c$, this means that
$\langle \Delta D \rangle / D = \Delta \phi / c^2$.  Thus, in this situation,
where curvature clearly plays an important role,  the redshifts can still be considered
to be `kinematic' in nature in the sense that equation (\ref{eq:stretchingrelation})
applies.   The question is whether this is a general phenomenon for  freely-falling observers.
If this were the case it would nicely
reconcile the elegant relativistic kinematic picture with the more pedestrian view of photons losing
energy climbing out of potential wells as well as with the the much maligned `expanding space' picture.

But while attractive, there are reasons to suspect that equation (\ref{eq:stretchingrelation})
is not, in fact, universal.  One is that it
does not even apply exactly for all pairs of observers in Minkowski space; though it
is a very good approximation for non-relativistic observers as the corrections are
of order $(v/c)^3$.  More importantly, the result above was for a very special form of the
potential and does not prove that this works out for more general potentials.
And indeed, for more realistic models for cluster haloes, 
the result would conflict with the conventional
view that, aside from complications arising from the transverse Doppler effect etc.,
the redshift is given by the usual static gravitational redshift.

The question of whether redshifts should generally be considered to be kinematic,
in the sense of  (\ref{eq:stretchingrelation}) 
(perhaps to an approximation with a precision growing in some calculable way
as the limit $v/c \rightarrow 0$ and/or small space-time curvature is approached), is readily answered by 
calculation.   In the next Section we compute the change in wavelength and change in
separation of a pair of observers who are freely falling in a weak, but otherwise arbitrary,
gravitational field.  We work to a precision sufficient to describe gravitational
redshifts (2nd order in velocity and 1st order in gravitational potential).
Consequently, the results are not applicable at high redshift, but should be
adequate to decide whether the the stretching law applies in the limit of
low redshifts and small curvature.

We find that, to the stated level of precision equation (\ref{eq:stretchingrelation}) 
applies for any emitter/receiver pair that are freely falling
in a {\em spatially constant tidal field\/}.  This is of slightly broader applicability
than just the redshift for a pair of FOs in homogeneous model, as it applies
also to non-fundamental observers (i.e.\ observers who may be moving with respect
to the local frame-of-rest defined by the matter).  
But in the presence of any inhomogeneity -- which implies spatially varying curvature --
the relationship does not hold beyond first order in the relative velocity. 
This is because the change in separation involves the gravity at the end points -- which
is the line integral of the tide -- but the wavelength change involves the integral of the gravity.
Only for a constant tide are these equal.
 If the {\em change\/} in the tide or curvature is small over the photon path length then  (\ref{eq:stretchingrelation})  may be
a very good approximation, but in general, applying this  gives errors that can be as large as the
gravitational redshift.  Indeed, in astronomical situations in which an unsophisticated
astronomer or physicist would describe the redshift as essentially gravitational in nature, 
the `velocity' that, in the Doppler formula, gives the redshift tends to be
very different from the real velocity.  

To show this it is sufficient to use simple special relativity and gravitational
redshifts as Einstein would predict from Newtonian gravity.  This is 
adequate for systems of relative velocity substantially less than $c$. 
We also work in terms of physical velocities and positions, which, we
believe, helps clarify what is going on.  

\section{Analysis}

To calculate changes in the wavelength of light exchanged, or
the distance, between a pair of freely falling particles (an emitter `1'
and a receiver `2') we imagine a family of non-inertial
observers who lie at the grid-points of  a non-rotating (as determined by gyroscopes) and non-expanding rigid lattice
armed with clocks and rulers to determine rates of motion of observers in
their vicinity and weighing scales
to determine their acceleration $\ba$, which is just the reflex of the Newtonian gravity $\bg$.  

Alternatively one could imagine a fleet of rocketeers who adjust their
thrusters to maintain a rigid, and non-rotating, spatial relationship with their neighbours and who report the
value of the acceleration of test particles they release; this would provide the Newtonian gravity $\bg(\br)$ relative
to one of their number -- the `reference observer' -- who is arbitrarily chosen and who
does not activate his or her thrusters.  The absolute value of the acceleration is somewhat arbitrary,
but differentiating $\bg(\br)$ gives, unambiguously, the tidal field.  

In either case, if a pair of these non-inertial observers exchange a photon, the redshift is just $\Delta \lambda / \lambda = - \int d\br \cdot \bg / c^2$.


\begin{figure}
\setlength{\unitlength}{2.100000cm}
\begin{picture}(4,2.6)(-2,-1.3)
\thinlines
\put(0.0,-1.3){\vector(0,1){2.600000}}
\put(0.100000,1.200000){\makebox(0,0){y}}
\put(-2,0.0){\vector(1,0){4.000000}}
\put(1.900000,0.100000){\makebox(0,0){x}}
\put(1.400000,0.300000){\vector(0,-1){1.300000}}
\put(1.400000,0.400000){\makebox(0,0){$\Delta y$}}
\put(1.400000,0.500000){\vector(0,1){0.500000}}
\multiput(-1.950000,-1.000000)(0.200000,0){20}{\line(1,0){0.100000}}
\multiput(-1.950000,1.000000)(0.200000,0){20}{\line(1,0){0.100000}}
\thicklines
\qbezier(-1.2, -1.0)(0,0)(1.2,1.0)
\thicklines
\put(-1.200000,-1.000000){\vector(-1,0){0.700000}}
\thinlines
\qbezier(-1.900000,-1.000000)(-0.350000,0.000000)(1.200000,1.000000)
\put(-1.550000,-0.900000){\makebox(0,0){$-\beta\hat{\bx}$}}
\put(-1.200000,-1.000000){\circle{0.030000}}
\put(-1.200000,-1.150000){\makebox(0,0){E$=r_1$}}
\put(-1.900000,-1.000000){\circle{0.030000}}
\put(-1.900000,-1.150000){\makebox(0,0){E'}}
\thicklines
\put(0.500000,1.000000){\vector(1,0){0.700000}}
\thinlines
\qbezier(-1.200000,-1.000000)(-0.350000,0.000000)(0.500000,1.000000)
\put(0.850000,1.100000){\makebox(0,0){$\beta\hat{\bx}$}}
\put(1.200000,1.000000){\circle{0.030000}}
\put(1.200000,1.150000){\makebox(0,0){R$=r_2$}}
\put(0.500000,1.000000){\circle{0.030000}}
\put(0.500000,1.150000){\makebox(0,0){R'}}
\qbezier(0.600000,0.000000)(0.600000,0.250000)(0.480000,0.400000)
\put(0.4,0.15){\makebox(0,0){$\theta$}}
\put(-1.1,-0.3){\makebox(0,0){$D_R$}}
\put(0.15,0.8){\makebox(0,0){$D_E$}}
\end{picture}
\caption{Dashed lines are paths of emitter and receiver, in flat space-time and in the centre of velocity frame,
who exchange a photon (heavy line)
that travels in the $x,y$ plane at angle $\theta$ to the $x$-axis (which we can chose to be aligned with the 
velocity difference) from point E $=\br_1$ to point R $=\br_2$.  
\comment{have emphasised correspondence of (E,R) to (1,2).}
Arrows indicate the velocity ($\beta = v/c$).
The location of the emitter and receiver at the times of 
reception and emission respectively are E' and R'.
Distances between the observers in centre of velocity frame at
times of emission ($D_{\rm E} = {\rm E R'}$) and reception ($D_{\rm R} = {\rm RE'}$) are indicated by thin lines and
are given by $\Delta y \sqrt{1 + \beta^2 \pm 2 \mu \beta} / \sin(\theta)$, with $\mu \equiv \cos(\theta)$.}
\label{fig:flatspacexyfig}
\end{figure}
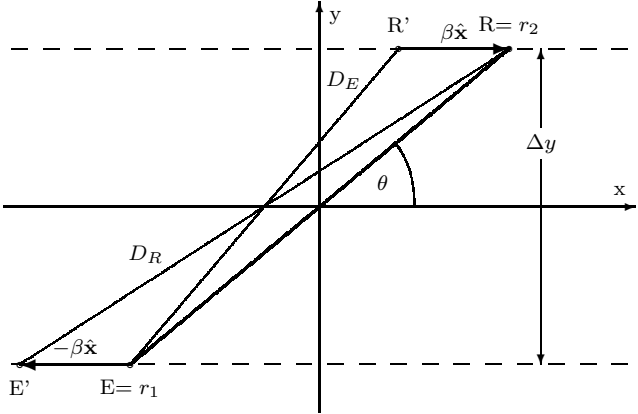

First consider a pair of observers in empty space.  Let us work in the centre of velocity frame -- the `lab' frame -- and, without loss of generality,
let the observers be moving in the $x,y$ plane with velocities along the $x$-axis $\bv = \pm \beta \hat{\bx} c$ with $\beta =  |\bv_2 - \bv_1| / 2 c$ 
at some distance $\pm \Delta y / 2$ from the $x$-axis.  Let them exchange a photon that, in the 
lab-frame has 4-momentum $p = \{p_0, \bp\} = \{p_0, p_x, p_y, 0\}$ and define the angle $\theta = \tan^{-1}(p_y/p_x)$ (see Fig.\ \ref{fig:flatspacexyfig}).  
Boosting this into the frames of the observers gives $p_0' = p_0 \gamma (1 \pm \mu \beta)$ where \comment{added:} $\mu = \cos(\theta)$ and, as usual, 
$\gamma = (1 - \beta^2)^{-1/2}$.
Defining $\lambda$ and $\Delta \lambda$ such that $\lambda_{\rm em} = \lambda - \Delta \lambda / 2$ and
$\lambda_{\rm rec} = \lambda + \Delta \lambda / 2$ then 
$\Delta \lambda / \lambda = p_x \Delta v /  |\bp| c = \bn \cdot (\bv_2 - \bv_1) / c = 2 \mu \beta$
where $\bn = \bp / |\bp|$.  Note that $\bn$ is not exactly the same as the photon directions as perceived by the observers as
these will be aberrated, with changes of angle that are first order in $v/c$.  Note also that there is no transverse Doppler
between the observers (as required by symmetry) though an observer in the lab frame
would see the transverse Doppler shift.  Had we defined the wavelength shift in the more conventional way
as $\Delta \lambda / \lambda = (\lambda_{\rm rec} - \lambda_{\rm em}) / \lambda_{\rm em}$ then there
would be additional second order terms.  The definition used here leads to a cleaner result, and is legitimate as
we will define $\Delta D$ in the same way below.

If we now switch on gravity, then the emitter and receiver velocities will become time dependent, and their paths, as well as those of
the photons, will become very slightly bent.  
The rocketeer's servo-controlled thrusters will fire, and they will sense their
acceleration and will also start to perceive that their
clocks start to slowly drift out of synchronicity with their neighbours 
(by an amount that they can correct for if they wished using their measurable non-gravitational acceleration).  
The redshift is now the product of 3 terms; local Lorentz boosts to or from the emitter and receiver's frame
into the frame of the local rocketeer and then a static gravitational redshift as the
photon propagates between the two rocketeers.  The Lorentz boosts will now no longer be
perfectly symmetric, but the difference between the Lorentz $\gamma$ factors is $\sim \delta \gamma \sim v \delta v / c^2$,
but with $\delta v \sim g \Delta t \sim \phi / c$ this is of third order in $v/c$ and so we can ignore it.
Similarly, any bending angles are on the order of the potential divided by $c^2$, so any change to the fractional wavelength shift which is already of first order in $v/c$ is also third order and can be neglected, as can any corrections from
the drifting of the clock synchronization.  The only effects that appear
at our stated precision goal is the change in the first order Doppler effect and the
static gravitational redshift:
\begin{equation}
\begin{split}
\frac{\Delta \lambda}{\lambda} & = \bn \cdot (\bv_2(t_2) - \bv_1(t_1)) / c
- \int\limits_{\br_1}^{\br_2} d\br \cdot \bg(\br)  / c^2 \\
& = \bn \cdot (\bv_2 - \bv_1)_{t_1} / c + \int\limits_{\br_1}^{\br_2} d\br \cdot (\bg_2 - \bg(\br)) / c^2
\label{eq:dloglambda}
\end{split}
\end{equation}
Thus, by working in a frame which is close to the centre-of-velocity frame we end up
with a simple Newtonian looking result.

If, as in the second line, we decompose the redshift into a first-order
Doppler or kinematic component and a gravitational redshift then 
only changes in the gravity vector with position -- to lowest order
the tidal field -- appear in the latter, consistent with the idea that
one can `transform away' any constant gravitational acceleration.
The asymmetry in this formula -- the fact that $\bg_2$ appears in the integral -- is 
a consequence of the fact that we have chosen to use the first order Doppler
term at the initial time $t_1$.  Had we used the final time then we would
have $\bg_1$ in the integral and if we had used the relative velocity
at the time the photon is half way along its path -- arguably the most
natural choice --  we would have $(\bg_1 + \bg_2)/2$ in the integral.

However, had we tried to decompose $\Delta \lambda / \lambda$ into a first order
effect from the velocity difference at {\em different times\/} $\bv_2(t_2) - \bv_1(t_1)$ 
as in the first line of (\ref{eq:dloglambda}) then the
gravitational effect is $-\int d\br \cdot \bg / c^2$, which is
a poorly defined concept in Newtonian gravity.  It is
not measurable from local geodesic deviation measurements;
it may in principle have a contribution from structures at arbitrary large distances;
as discussed, its absolute value is arbitrary.   But there
is no real physical problem here; the combination of these individually
poorly defined terms does not depend on the arbitrary choice of
zero-point of $\bg$.  But it is salutary, nonetheless, that any difference
of velocities at different times is poorly-defined in Newtonian gravity; since
GR contains Newtonian gravity as a limiting case one should also be wary
about any discussion, or calculation, invoking difference of velocities
at different times.

This result resolves the issue raised in the introduction 
which was the legitimacy of `transforming away'  the gravity in the vicinity of the
receiver.  This, in effect, is what we have done in the second line of (\ref{eq:dloglambda})
where the integrand is zero at the location of the receiver (particle 2) and consequently 
the gravitational term will, for a smooth potential, grow quadratically with
distance from the receiver.  That is fine, but only gives the correct redshift if it is combined with the
velocity of the receiver determined at the time of emission.  For randomly chosen observers
in dynamical equilibrium the average velocity at the emission time is not
zero, and the net result is the usual static gravitational redshift (though there is
still the complication of the traverse Doppler effect if one works in the
rest-frame of the cluster).

Regardless of which velocity is chosen to define the `kinematic' component, the
gravitational component does involve the tidal field.  The only distinction
is whether the gravitational component is fully determined by the tide, as it is
if the relative velocity is taken to be in the rest-frame, or whether it also involves
the gravity, as is the case if the velocity difference is at different times.  This
is at odds with Synge's statement that the curvature does not appear in the redshift.

What about the change in the separation during the light-propagational time?
Letting the centre of velocity frame separations, in the absence of gravity, at
reception and emission be $D_R,D_E = (D \pm \Delta D / 2)$ we see
from the caption of Fig.\ \ref{fig:flatspacexyfig} that $\Delta D / D$ is not precisely
the same as the flat-space $\Delta \lambda / \lambda = 2 \mu \beta$.
But the difference is of order $\beta^3$, so to our precision goal we
can take them to be equal.  Switching on gravity, the fractional change of the separation between the particles as
measured by lattice based observers -- i.e.\ the change in the proper
separation in the centre of velocity frame -- is
\begin{equation}
\Delta D/D = \bn \cdot (\bv_2 - \bv_1)_{t_1} / c + \Delta {\br} \cdot (\bg_2 - \bg_1)  / 2  c^2 
\label{eq:dlogD}
\end{equation}
which is also a simple Newtonian looking result.
Unsurprisingly, to first order in the relative velocity the fractional changes in wavelength
and separation are identical.  Both contain an additional gravitational term
that is, to lowest order, a tidal effect.  In general, these gravitational
effects are not precisely equal -- $\bg_2 - \bg_1$ is the integral of the tide while
the wavelength shift involves the integral of the gravity -- so the `cosmological relation' that wavelengths vary precisely
in proportion to the source-observer proper separation, does not hold in general.

But if the tide is spatially constant -- i.e.\ the potential has no spatial derivatives higher than second --
then the gravity varies linearly with position then we can write $\bg(\br) = \bg_1 + (\bg_2 - \bg_1) |\br - \br_1| / |\br_2 - \br_1|$
and the gravitational terms are readily found to be identical.
Thus the relation seen in cosmology is of wider generality, and applies for an arbitrary pair of particles
moving in a field that has a spatially constant tide.   This includes,
as a special case, a pair of particles with relative motion along their
separation in a quadratic potential as in a FRW model containing matter
and/or dark energy.   Note that there is no need for the particles to be
co-moving with the matter density, though again this result does
apply in that situation.  

This is one of the two main results of this paper: a constant tide stretches wavelength of radiation
just as it changes the separation of test-particles.  Arguably this `explains' the apparent
stretching of wavelength of light by the expansion of space in FRW models.

\begin{figure}
\begin{picture}(238,238)(-119,0)
\thinlines
\put(-119,63.466667){\vector(1,0){238}}
\put(111.000000,58.466667){\makebox(0,0){$r$}}
\put(-79.333333,61.483333){\line(0,1){3.966667}}
\put(79.333333,61.483333){\line(0,1){3.966667}}
\put(0,3.966667){\vector(0,1){71.400000}}
\put(55.533333,71.400000){\makebox(0,0){$W_2=(r^2-d^2)\theta(|r|-d)/2$}}
\put(-119,119.000000){\vector(1,0){238}}
\put(111.000000,114.000000){\makebox(0,0){$r$}}
\put(-79.333333,117.016667){\line(0,1){3.966667}}
\put(79.333333,117.016667){\line(0,1){3.966667}}
\put(0,83.300000){\vector(0,1){71.400000}}
\put(35.700000,150.733333){\makebox(0,0){$W_1(r)=-W'_2(r)$}}
\put(-119,214.200000){\vector(1,0){238}}
\put(111.000000,209.200000){\makebox(0,0){$r$}}
\put(-79.333333,212.216667){\line(0,1){3.966667}}
\put(79.333333,212.216667){\line(0,1){3.966667}}
\put(0,162.633333){\vector(0,1){71.400000}}
\put(35.700000,230.066667){\makebox(0,0){$W_0(r)=-W'_1(r)$}}
\thicklines
\put(-75.366667,214.200000){\line(0,-1){47.600000}}
\put(-83.300000,214.200000){\line(0,-1){47.600000}}
\put(-83.300000,166.600000){\line(1,0){7.933333}}
\put(-83.300000,214.200000){\line(-1,0){35.700000}}
\put(83.300000,214.200000){\line(0,-1){47.600000}}
\put(75.366667,214.200000){\line(0,-1){47.600000}}
\put(75.366667,166.600000){\line(1,0){7.933333}}
\put(83.300000,214.200000){\line(1,0){35.700000}}
\put(75.366667,214.200000){\line(0,1){7.933333}}
\put(-75.366667,214.200000){\line(0,1){7.933333}}
\put(-75.366667,222.133333){\line(1,0){150.733333}}
\qbezier(0.000000,119.000000)(39.666667,101.150000)(79.333333,83.300000)
\put(79.333333,119.000000){\line(0,-1){35.700000}}
\put(79.333333,119.000000){\line(1,0){39.666667}}
\put(79.333333,126.000000){\makebox(0,0){$+d$}}
\qbezier(0.000000,119.000000)(-39.666667,136.850000)(-79.333333,154.700000)
\put(-79.333333,119.000000){\line(0,1){35.700000}}
\put(-79.333333,119.000000){\line(-1,0){39.666667}}
\put(-79.333333,112.000000){\makebox(0,0){$-d$}}
\qbezier(0.000000,15.866667)(-39.666667,15.866667)(-79.333333,63.466667)
\put(79.333333,63.466667){\line(1,0){39.666667}}
\qbezier(0.000000,15.866667)(39.666667,15.866667)(79.333333,63.466667)
\put(-79.333333,63.466667){\line(-1,0){39.666667}}
\end{picture}
\caption{The upper plot shows, schematically, the dimensionless
weight function $W_0(r ) = \theta_+(r) \theta_-(r) - d(\delta(r-d) + \delta(r+d)) / 2$
that, when multiplied by the gravity $\phi'(r )$ gives the difference
$\Delta \lambda / \lambda - \Delta D / D$.
The Dirac $\delta$-functions are shown as the narrow box-cars at
$r = \pm d$ and together have (minus) the same weight as the central box-car.
As described in the text, this difference can also be computed
as a weighted average of the tide $\phi''(r )$ using the
weighting function $W_1(r )$ shown in the centre plot, which
is (minus) the integral of  $W_0(r )$, and also has zero net weight.  The third way to compute
the difference is averaging the gradient of the tide $\phi'''(r )$
with the weight function shown in the bottom plot.
}
\label{fig:kernels}
\end{figure}
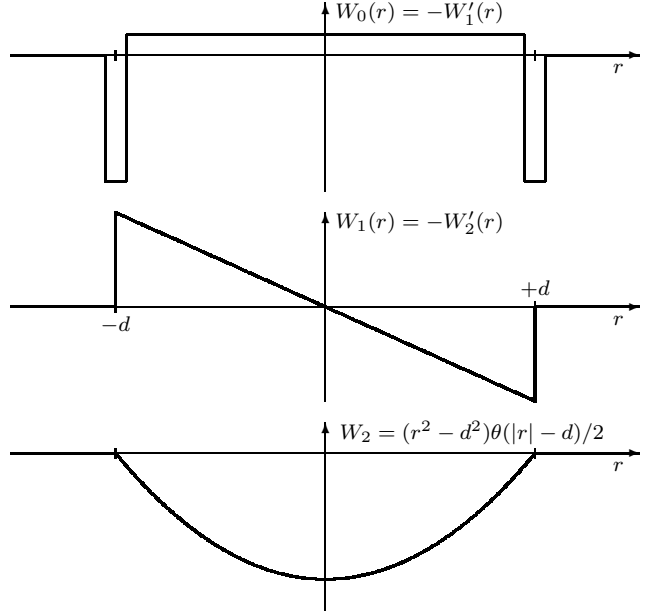

To highlight the differences between separation and wavelength changes -- 
what one might call the `non-kinematic' component of the redshift --
and to see how this depends on the tide (and its derivatives) we note the
following:  

First, we can write 
\begin{equation}
\Delta D / D = \bn \cdot (\bv_2 - \bv_1)_{t_1} / c - \frac{\Delta r}{2c^2} \int dr \phi''(r)
\label{eq:dlogDfromtide}
\end{equation}
where $\phi(r) = \phi(r \bn)$ is the gravitational potential and prime denotes the operator $\partial_r = \bn \cdot \nabla$,
i.e.\ the spatial derivative along the photon path,
so $\phi'(r) = \bn \cdot \nabla \phi(\br) = - \bn \cdot \bg$. 
Thus the gravitational contribution to $\Delta D / D$ is the average of the tide along the photon path
times $(\Delta r / c)^2 /2$.

Second, taking the difference of (\ref{eq:dloglambda}) and (\ref{eq:dlogD}), we have
\begin{equation}
\Delta\log(\lambda / D)= \frac{1}{c^2}
\left(d\  \bn  \cdot (\bg_1 + \bg_2) -  \int d\br \cdot \bg \right)
\label{eq:Dlnlambda-DlnD}
\end{equation}
where $d \equiv |\br_2 - \br_1| / 2$.
Taking the origin of coordinates to lie at $(\br_1 + \br_2) / 2$ for simplicity,
this is a weighted average of the gravity $-\phi'$:
\begin{equation}
\Delta\log(\lambda / D)= \frac{1}{c^2} \int dr\ W_0(r ) \phi'(r ) 
\label{eq:Dlnlambda-DlnD0}
\end{equation}
with dimensionless weighting function 
\deltext{$W_0(r) \equiv \theta_+(r) \theta_-(r) - d(\delta(r-d) + \delta(r+d)) / 2$;}
\newtext{$W_0(r) \equiv \theta_+(r) \theta_-(r) - d(\delta(r-d) + \delta(r+d))$,}
where $\phi(r) \equiv \phi(\br = r \bn)$; 
and where $\theta_\pm(r) \equiv \theta(\pm r - d)$
with $\theta(r )$ and $\delta(r )$ denoting the Heaviside function
and the Dirac delta function respectively.
The weight function $W_0(r )$ is shown schematically as the upper
plot in Fig.\ \ref{fig:kernels}.  The product of Heaviside functions is zero for $|r| > d$ so the
range of integration is now unrestricted.  The integral of $W_0(r )$ over all $r$ vanishes, so
we can immediately integrate by parts to eliminate the gravity and write (\ref{eq:Dlnlambda-DlnD0})
as an integral of the tide:
\begin{equation}
\Delta\log(\lambda / D)= \frac{1}{c^2}   \int dr \; W_1(r ) \phi''(r )
\label{eq:Dlnlambda-DlnD1}
\end{equation}
where $W_1(r ) = - \int dr\; W_0(r ) = - r \theta_+(r ) \theta_-(r )$ which is shown is the middle plot in
Fig.\ (\ref{fig:kernels}).  But the integral of $W_1(r ) $ also vanishes (so, as already mentioned
for spatially constant tide $\Delta \lambda / \lambda = \Delta D / D$)
and we can integrate once more by parts to express the difference of these as a
weighted average of $\phi'''(r )$:
\begin{equation}
\Delta\log(\lambda / D)= \frac{1}{c^2}  \int dr \; W_2(r ) \phi'''(r )
\label{eq:Dlnlambda-DlnD2}
\end{equation}
with weight function $W_2(r ) = (r^2 - d^2)  \theta_+(r ) \theta_-(r ) / 2 = (r^2 - d^2) \theta(|r| - d) / 2$ now shown as the bottom plot in figure (\ref{fig:kernels}).
This is the second main result of this paper.

\section{Discussion}

We have made various simplifying assumptions.  The results are
only valid up to 2nd order in velocity and 1st order in the potential, but this is
adequate to encompass the phenomena that have proved controversial.
We have ignored any time variation of the potential, and we have
also ignored any bending of light rays.  These are higher order effects;
the Rees-Sciama (1968) effect is of order $(v/c)^3$ for instance.
We can also ignore spatial curvature.  In \deltext{a} \newtext{an} FRW model, for example, the distance
between two  FOs as measured by observers on a ruler is not precisely the same
as the distance as determined by a chain of FOs lying along a geodesic in the 3-space of constant
proper time since the big-bang, but the fractional difference is at most of order the
square of the separation in units of the curvature radius (or of order $(v/c)^2$) so any
effect on $\Delta D / D$ is of higher order.  
Similarly the asynchronicity between clocks carried by our non-inertial grid-based
observers is negligible at our stated level of precision. 

More fundamentally
while matter tells \deltext{space-{\em time\/}} \newtext{{\em space-time\/}} how to curve, it is
only the curvature of time that tells non-relativistic matter how to move, and is is
also only the time-time part of the metric perturbation that is relevant for the
calculation of the redshift.   So at the specified level of precision we
can ignore the spatial part of the metric.

We have asked: what is the domain of validity of the relationship between
wavelengths and emitter/receiver proper separation (\ref{eq:stretchingrelation}) that we see for
FOs in homogeneous models?
We have shown that a spatially constant tide stretches
wavelength in exactly the same way it affects the observers' separation, but if the tide varies
with position the relationship between wavelength and separation is modified.

From this perspective, the perfect correlation seen between changes in
wavelengths $\Delta \lambda / \lambda$ of light exchanged between FRW FOs and the change $\Delta D / D$ in the
space (between said FOs) is not a causal relationship, rather both the change in the wavelength and the change in the
space between the observers are `caused', or determined, by a combination of the observers' initial velocities and the tidal
field in which they and the photons propagate.   Echoing Whiting (2004),
the expansion rate defined by the matter content of the universe is
irrelevant (which is a jolly good thing if the universe has a cosmological 
constant or a scalar field to realise dark energy since neither defines
either a frame of motion or a state of expansion).  Rather, on the parabolic potential generated by gravitating
matter and dark energy one can have emitter/receiver
pairs that recede from each other or pairs that approach each other,
and what determines both the changes in the wavelengths and 
proper separations is a combination of initial conditions and the curvature, or tidal field.

The perfect correlation of $\Delta \log(\lambda)$  and  $\Delta \log (D)$ in homogeneous models 
can be considered to be a reflection of the
symmetry of the gravitational fields that are allowed in these models, in accord with the conjecture of Melia (2012).

In the Introduction we asked what redshift would be seen by a
pair of observers in a cluster who have the same separation at emission as at reception.
Our analysis shows that they do {\em not\/} see the effect of any local tidal field
\newtext{and if they were residing in a constant density cluster core they would
see no redshift}.  \comment{Following needed correcting.} \deltext{What happens is
that any gravitational stretching of the wavelength that would be perceived
in the frame of the rigid non-inertial observers is counteracted by the
red- or blue-shifts in boosting from the observers' frames to the rigid frame and back again (if, for example they were
moving apart at the emission time they will be moving together again by the time of reception).}
\newtext{From the perspective of their centre of mass, the observers were moving
apart at the moment of emission, but falling back together by the time of reception,
so the Doppler shifts would cancel, and the net gravitational redshift also vanishes as any energy
gained by the photon on the first half of its journey is cancelled by the redshift on the second half.
The receiver could justifiably consider itself to be at rest at the centre of the parabolic potential well
generated by the uniform matter density.  From that  perspective the net redshift
vanishes because the Doppler redshift at emission is cancelled by the gravitational
blue-shift as the photon rolls down the potential to the receiver.}

In an inhomogeneous system such as the solar system, or a galaxy, cluster or supercluster,
the tidal field necessarily varies with position.  There is then a non-kinematic component to 
the redshift that violates (\ref{eq:stretchingrelation}) and which is essentially gravitational in nature.
Combining this with the kinematic redshift component, if any, one obtains complete consistency
with the conventional view of the gravitational redshift in clusters of galaxies and other gravitating systems.
Note that our description of components of the redshift is different from the  terminology of Chodorowski (2011) who was considering the
gravtiational component of the redshift in FRW models that arises if the `kinematic' component is
taken to be the relative velocity at emission or reception rather than the average velocity. 
Here we consider redshifts between observers in FRW models to be purely kinematic in the sense that (\ref{eq:stretchingrelation}) is obeyed.

For emitter/receiver
pair separation that is small compared to the size of the gravitating system
the difference between the fractional change in the wavelength and
separation is on the order of the gravitational potential well depth times
the cube of the separation in units of the overall system size.  
This is seen most easily from (\ref{eq:Dlnlambda-DlnD2}), and the fact that
$W_2(r ) \sim d^2$, which together imply $\Delta\log(\lambda/D)  \sim (d/R)^3 \phi/c^2$ \newtext{where $R$ is the size of the system}.
So, just as the local {\em gravity\/} is invisible to freely-falling observers, as far
as the ratio of wavelength to separation is concerned, the local {\em tide\/} is
also invisible.
But if the path length
has a similar size to the entire system the error is on the order
of the gravitational potential.

\begin{figure}
\includegraphics[angle=-90,width=84mm]{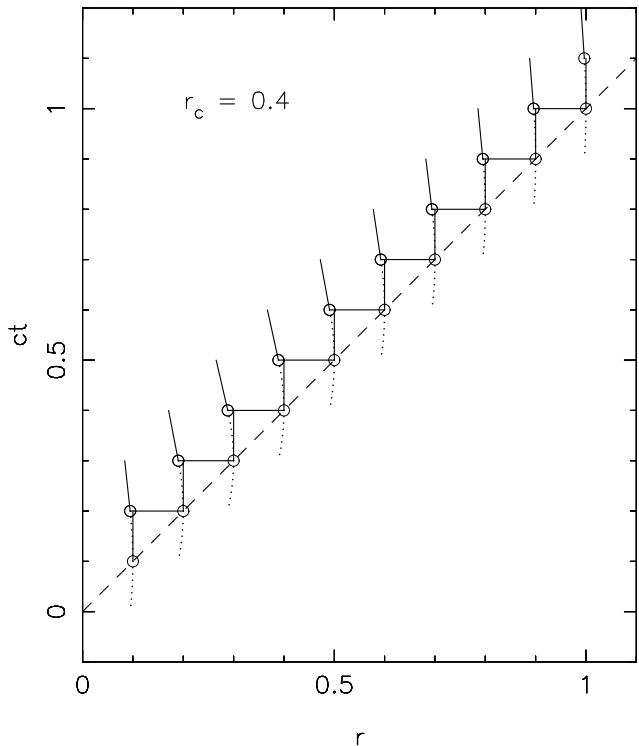}
\caption{Illustration of the example described in the text.  Here we have a potential
with gravity $g \sim r / (r^3 + r_c^3)$ which is like that for a uniform density
sphere at $r \ll r_c$ and is Keplerian at $r \gg r_c$.  
The redshift between an emitter at $r=0$
and a distant receiver at $R \gg r_c$ who happens to be turning
around at the instant of reception is
just the static gravitational redshift.  But following Synge \& Peebles the wavelength ratio is also equal
to the product of Doppler shifts between pairs of fictitious particles
along the photon path which, for simplicity, can be taken  to be all on
radial orbits (dotted lines) that happen also
to be turning around as the photon (dashed line) passes.
The velocities are supposed all to be small compared to $c$
but have been exaggerated here for clarity.
Now the pairwise velocity differences have to be calculated
in the rest-frame of each pair; i.e.\ the differences are between the space-time
points connected by the horizontal lines.  If the velocities were differenced
along a continuous path -- at a constant time say -- the sum of the pairwise
velocities and the relative velocity of the end points has to be equal.  But when
a null path is chopped up into a set of space-like intervals like this
the connection between the `relative velocity' obtained by summing pairwise differences and the
true relative velocity is broken. For this type of potential, the resulting
net `relative velocity'  is dominated by the transition region $r \sim r_c$, where
there are only fictitious particles.   For $R \gg r_c$ the true rate of change
of the separation of the only two real particles involved is much smaller than that
calculated by summing these fictitious velocity differences.}
\label{orbits_fig}
\end{figure}

To see better how this relates to Synge's result that the redshift
is always given by the Doppler formula, 
consider the case of an emitter at the centre of the potential
for a small uniform spherical distribution of matter 
of mass $M$ and radius $r$ and a receiver outside at distance $D \gg r$ 
who happens to be at rest at the moment of reception.
In this situation, the redshift is just the static gravitational redshift: $\Delta \lambda / \lambda = \int dr \; g / c^2 \sim G M / c^2 r$.
But the more distant the receiver, the smaller any fractional change in the
emitter/receiver proper separation during the time of flight: $\Delta D \lesssim (G M  / D^2) (\Delta t)^2 / 2$ which
implies $\Delta D/D \lesssim G M / c^2 D \ll \Delta \lambda / \lambda$.  Evidently
the kinematic relation does not apply here.  But, following Peebles,
we can still break the net wavelength ratio down into the product of
ratios between a set of pairs of neighbouring particles.  We can take these to be
particles on a set of radial orbits such that each
particle is at apogee at the time the photon passes (see Fig.\ \ref{orbits_fig}).  Thus the $n$th particle
has zero velocity as the photon passes it, as does the $(n+1)$th particle.
In the {\em rest-frame\/} of the pair of particles, however, at the time the photon passes the $(n+1)$th particle the
$n$th particle will have started to fall back towards the mass and will
have picked up a velocity $\delta v = g \delta t$.  So the Doppler shift,
evaluated using this small rest-frame velocity, is $\delta v / c = g \delta r / c^2$, which
is just the gravitational redshift for this  element of the path, and
this $\delta v$ is also the same as the result of parallel transporting the
4-velocity of the $n$th particle along the null ray and subtracting it from the
4-velocity of the $(n+1)$th particle.   Either way, integrating these
velocity increments gives the gravitational redshift, so there
is no mathematical conflict with Synge.    But this `velocity' is not related in any
sensible way to the rate of change of the emitter-receiver proper separation which is much smaller.
It is therefore misleading to say that redshifts in the situation described here
-- that a non-sophisticated physicist would say are essentially gravitational -- are kinematic in nature.


If instead we consider a similar pair of particles in a quadratic potential -- where, unlike the Keplerian example
above the tidal field is spatially constant -- the redshift is again just the gravitational redshift,
but in this case this is not inconsistent with the the kinematic interpretation
since at the time of emission the receiver was indeed closer to the 
source by an amount such that the fractional change in $\lambda$ is indeed the same as the fractional change in
separation.

We do not think that Synge would object strongly to our conclusions.  While he did say that if one
were to attribute a cause to the spectral shift one would have to say that it is caused by the
relative velocity of the source and observer and is not a gravitational effect, \deltext{which is perhaps unfortunate,} this should not be taken out of
context.  He followed that immediately by emphasising that this is true only given the specific
definition of relative velocity in terms of parallel transport; that one is not obliged to accept that
definition; and that arguments about whether a redshift is a gravitational or a velocity effect
are futile without any attempt to analyse the meanings of the terms employed.  Unfortunately he
did not expand much on the physical interpretation of his relative velocity.  That is
what we have tried to do here.
Later in his book he adopts a relatively conventional decomposition of spectral shifts into
a product of velocity and gravitational effects, remarking that the earlier formalism should
only be taken as `a matter of speaking'.  

Synge's reason for saying the redshift is not gravitational was that the Riemann tensor does
not appear in the formulae.  This was emphasised by Narlikar (1994) and underlies
Bunn \& Hogg's narrative.  But gravity obviously does play a role.
We believe that this has contributed to the confusion over the nature of redshifts.
We cannot know exactly what Synge had in mind, but perhaps he was referring to the
fact that the parallel transport operation depends on the {\em connection\/}, which
contains first derivatives of the metric, rather than the second derivatives that appear
in the curvature or tide.  As we have discussed, there is something a little disturbing 
about this since in the Newtonian limit
this corresponds to the gravity, which is somewhat poorly defined, and that carries over into the
more general theory.  But to do anything useful with a redshift, one needs to compare
it with something else -- like the relative velocity, for instance -- in which case the
difference tells us something physically meaningful about the gravitational field and
hence the mass distribution that generates it.  As we have seen, provided the
velocity differences are taken to be at the same time there is no ambiguity since the
connection itself does not then appear.  

In conclusion, we hope that the discussion and rather elementary analysis presented here makes clear
that saying that redshifts are an effect of the relative velocity is either
meaningless, as it is if there is no means at one's disposal other than redshift
to measure relative velocity, or, unless the tide happens to be constant,  it is false.  
In many situations, of course, the kinematic relationship is an excellent approximation, because in
most circumstances the first order velocity effect dominates over any gravitational
effects, which are generally of second order in the velocity. But as regards the
latter, redshifts cannot, in general, be regarded as kinematic in nature.

\section{Acknowledgements}

We gratefully acknowledge stimulating discussions on this subject with Shaun Cole and with numerous
fellows of the Cifar Cosmology and Gravitation programme, \newtext{helpful suggestions from the referee,} 
and, especially, \newtext{discussions} with
Michal Chodorowski who spotted an error in an early draft.


\begin{thebibliography}{}
\bibitem[\protect\citeauthoryear{Bondi}{1947}]{1947MNRAS.107..410B} Bondi H., 1947, MNRAS, 107, 410 
\bibitem[\protect\citeauthoryear{Bunn \& Hogg}{2009}]{2009AmJPh..77..688B} Bunn E.~F., Hogg D.~W., 2009, AmJPh, 77, 688
\bibitem[\protect\citeauthoryear{Cappi}{1995}]{1995A&A...301....6C} Cappi A., 1995, A\&A, 301, 6
\bibitem[\protect\citeauthoryear{Chodorowski}{2007}]{2007ONCP....4...15C} Chodorowski M.~J., 2007, ONCP, 4, 15 
\bibitem[\protect\citeauthoryear{Chodorowski}{2011}]{2011MNRAS.413..585C} Chodorowski M.~J., 2011, MNRAS, 413, 585
\bibitem[\protect\citeauthoryear{Darling}{2013}]{2013ApJ...777L..21D} Darling J., 2013, ApJ, 777, L21 
\bibitem[\protect\citeauthoryear{Gunn}{1988}]{1988ASPC....4..344G} Gunn J.~E., 1988, in The Extragalactic Distance Scale (ASP Conference Series) , 4, 344, ed. S.\ van den Bergh \& C.\ Pritchet. 
\bibitem[\protect\citeauthoryear{Harrison}{2000}]{Harrison2000book} Harrison E.~R., 2000, Cosmology: The Science of the Universe, (Cambridge University Press, Cambridge)
\bibitem[\protect\citeauthoryear{Kaiser}{2013}]{2013MNRAS.435.1278K} Kaiser N., 2013, MNRAS, 435, 1278 
\bibitem[\protect\citeauthoryear{Lineweaver \& Davis}{2005}]{2005SciAm.292c..36L} Lineweaver C.~H., Davis T.~M., 2005, SciAm, 292, 030000
\bibitem[\protect\citeauthoryear{Melia}{2012}]{2012MNRAS.422.1418M} Melia F., 2012, MNRAS, 422, 1418 
\bibitem[\protect\citeauthoryear{Narlikar}{1994}]{1994AmJPh..62..903N} Narlikar J.~V., 1994, AmJPh, 62, 903 
\bibitem[\protect\citeauthoryear{Peacock}{2008}]{2008arXiv0809.4573P} Peacock J.~A., 2008, arXiv, arXiv:0809.4573 A diatribe on expanding space
\bibitem[\protect\citeauthoryear{Peebles}{1971}]{PeeblesPhysicalCosmology} Peebles P.~J.~E., 1971, Physical Cosmology, (Princeton University Press, Princeton)
\bibitem[\protect\citeauthoryear{Pound \& Rebka}{1959}]{Pound+Rebka59} Pound, R. V.; Rebka Jr. G. A., 1959, Physical Review Letters 3 (9): 439Ð441
\bibitem[\protect\citeauthoryear{Rees \& Sciama}{1968}]{1968Natur.217..511R} Rees M.~J., Sciama D.~W., 1968, Nature, 217, 511 
\bibitem[\protect\citeauthoryear{Rindler}{1977}]{1977book} Rindler W., 1977, Essential Relativity.  (Springer-verlag, New York)
\bibitem[\protect\citeauthoryear{Synge}{1960}]{1960grbook} Synge, J.~L., 1960, Relativity: the General Theory. (North-Holland, Amsterdam)
\bibitem[\protect\citeauthoryear{Whiting}{2004}]{2004Obs...124..174W} Whiting A.~B., 2004, Obs, 124, 174
\bibitem[\protect\citeauthoryear{Wojtak, Hansen, \& Hjorth}{2011}]{WHH} Wojtak R., Hansen S.~H., Hjorth J., 2011, Nature, 477, 567
\bibitem[\protect\citeauthoryear{Zhao, Peacock, \& Li}{2013}]{2013PhRvD..88d3013Z} Zhao H., Peacock J.~A., Li B., 2013, PhRvD, 88, 043013 
\end{thebibliography}
\end{document}